\documentclass{article}

\usepackage{epsfig}
\usepackage{graphics}
\usepackage{graphicx}

\title{Modelling axon growing using CTRW \thanks{The work
is partially supported by RAS-CNRS grant EDC26091 and RFBR grants 14-01-00379 and 13-01-12410}}

\author{Xavier Descombes, \\
{\it Inria Sophia Antipolis, France}\\
\\
Elena Zhizhina, Sergey Komech\\
{\it Institute for Information Transmission Problems,}\\
{\it Russian Academy of Sciences, Russia}
}

\date{}
\graphicspath{{./images-tex/}}

\begin{document}

\maketitle

\section{Description and analysis of the model}
The main goal of this study is to propose a mathematical model describing paths of the axon growth cones and differences in the behavior of normal and mutant axons. We introduce a probabilistic model for axon growing, such that each family of axons is described as an ensemble of trajectories of a continuous time random walk (CTRW) model under different parameters in the case of normal and mutant axons. We describe different regimes in the model and conclude how the behavior of axons depends on the parameters of the model. Biological observations of the axonal growth process say us that the guiding development of axons to their targets is operated by chemical signals from the cellular environment. To simulate this control mechanism we propose the CTRW model, where a random waiting time reflects a reaction time of the growth cones on the neighboring chemical environment.

Continuous-time random walks (CTRW) are a natural generalization
of the usual random walk. The mathematical analysis   can be dated back to
the pioneering work of   Montroll and Weiss in the sixties \cite
{MW65}. At present they are extensively used in applications to
physics, chemistry, and other sciences, see e.g. \cite{Kol09},  \cite{RG}.

We consider three families of (static) trajectories associated with axon growth cones. One of the family is for normal axons, the other two are for two different axon mutations. We don't have dynamical data, and any trajectory presents positions of an axon tip during the observation time. The observation time is the same for all axons, but we don't have information about the stopping time for development of each axon. On figures 1-3 we present three sets of trajectories of axon growth cones. In fact, axons on real data images have different starting positions, but we shifted all of them to have the same starting point.

\begin{figure}[h]
  \centering
  \includegraphics[height=4cm]{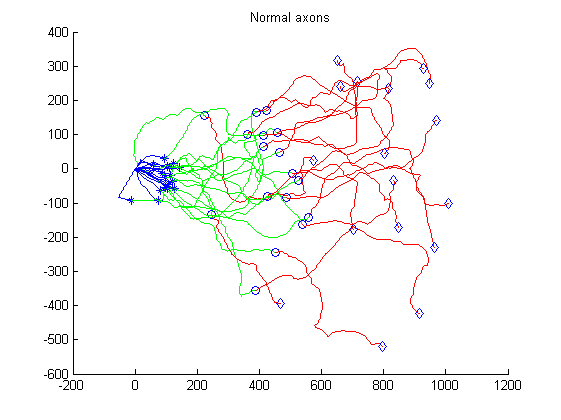}
  \includegraphics[height=4cm]{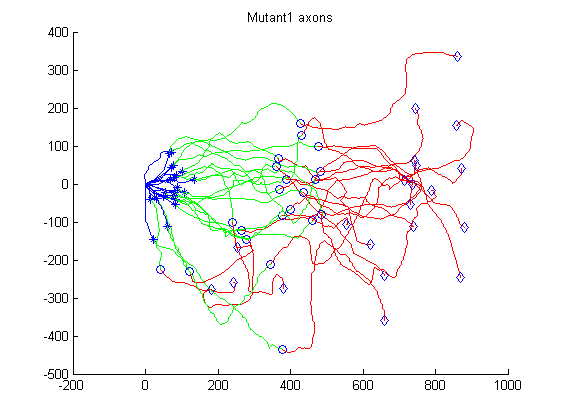}
  \includegraphics[height=4cm]{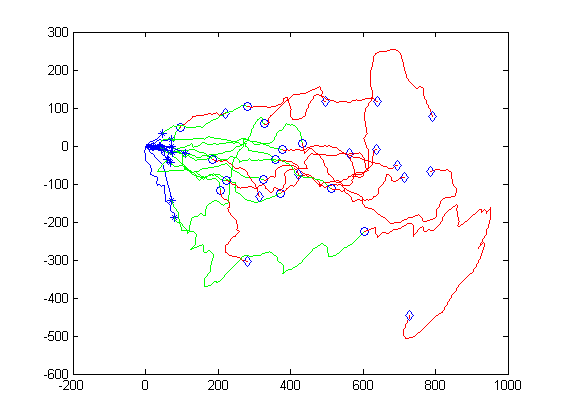}
  \caption{Axonal trajectories}
\end{figure}

\begin{figure}[h]
  \centering
  \includegraphics[height=4.5cm]{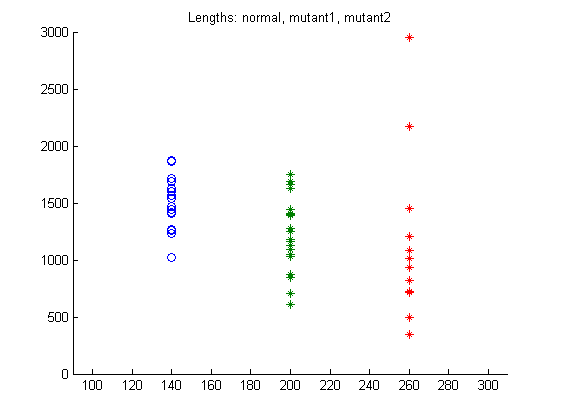}
  \caption{Length distributions}
\end{figure}

On fig. 4 we present the values of the lengths for the normal and mutant axons. One can see that the average length for the mutant axons is less then for the normal ones.

The second observation concerns the value of the average deviation of trajectories from the origin. In the case of mutant 2 trajectories it is much less then in the normal case. To explain these facts we propose  a model of axon growing based on a CTRW model. A CTRW model is a symmetrical random walk subordinated to a renewal process. It is defined by two probability distributions: a distribution $p(x \to y), \; x,y \in Z^d$ for a spacial random walk and a distribution $f(t), \; t \ge 0$ for a waiting time. Then CTRW can be described as follows: a particle wait at position $x$  a random time distributed by $f(t)$, then it jumps from $x$ to $y$ according to the distribution $p(x-y)$, and so on. Then a more long waiting time implies reducing of trajectories lengths.

Since we don't have any dynamical data for axon growing, we couldn't control the absolute time and should introduce in our model an artificial time. We study here three cases.
\\
{\it The first case (A uniform partition)}. Assuming an equal absolute time $T$ for growing of all axons and the existence of the average speed for any trajectory of the CTRW $X(t) \in Z^d,\, t \in [0,T]$ (depending on the length of the trajectory) we can consider uniform partitions of trajectories proportional to the length $L$ of axons: $X(0)=0, \; X(T) = x_{end}, \; X (\frac{k}{10} T) = x_k, \; k=0,1, \ldots, 10$, where $x_{end}$ is the position of the end of the trajectory of the length $L$, and $x_k, \; k=0, 1, \ldots, 10$ are points on the trajectory corresponding to the length  $\frac{k}{10} L, \; k=0, 1, \ldots, 10$. \\
{\it The second case (A uniform partition after elongation)}. In this case we assume different time for axon growing but the same evolving time $T$ for all axons. For long axons we consider again an equal absolute growing/evolving time as above, assuming different average speeds of the growth dependent on the length $L$. For short axons we say that the growing process stops at some moment of time, and after that the short axons just keep their final position in evolution process. Let $L_n$ be an average length of trajectories of normal axons, then we put $L_n=T$ as the absolute evolving time and consider the following elongation in time for all short axons with a length $L$ less than $L_n=T$: we take the last part of the trajectory (from "time" $L$ to time $T$) coinciding with the end point of axons. Thus we obtain that all short trajectories evolve during the same time $T=L_n$ with a constant speed 1, and we assume that it is the same absolute growing/evolving time $T=L_n$ for long trajectories. Finally, we construct the uniform partitions depending on the length of axons $L$, which is the same as in the first case. \\
{\it The third case (A non uniform partition)}.  Assuming again the same absolute time for growing of all axons and using observations from biological experiments that a speed of axon growing is decreasing in time (especially for mutant axons), we consider a non uniform partitions of trajectories corresponding to the following lengths $l(k)$:
$$
l(k) \ = \  \frac{\alpha L}{1-e^{-\alpha}} \int_{0}^{t_k} e^{-\alpha t} \ dt, \quad t_k = \frac{k}{10}, \quad k = 0,1, \ldots 10
$$
with $\alpha = 2$.
\\

Since we have a permanent drift along the X-axis, then we will consider the deviation only along the Y-axis, considering X direction as a direction of an artificial time. We will use the following result for the mean squared deviation $\langle Y^2(t) \rangle$, where $Y(t)$ is the position of the Y-coordinate of the CTRW when time $t$ is large enough.
\\

{\bf Proposition.} {\it Let $p(u) = p(-u)$ is a distribution of a symmetrical random walk on $Z^1$, and we consider two cases:\\
1) diffusive behaviour: $ \langle \tau \rangle  \ = \ \int_0^{\infty} t f(t) dt \ < \ \infty,$ \\
2) anomalous diffusion: $ \langle \tau \rangle =  \ \infty$  and $ f(t) \ \sim \ \frac{1}{t^{\alpha + 1}}$ as $t \to \infty$ with $\alpha \in (0,1)$.

Then for large enough $t$ we have for the mean squared displacement
\begin{equation}\label{1}
\langle Y^2(t) \rangle \ \approx \ \frac{2a}{\langle \tau \rangle} \ t \quad \mbox{in the first case}
\end{equation}
and
\begin{equation}\label{2}
\langle Y^2(t) \rangle \ \approx  \  t^{\alpha} \quad \mbox{in the second case with} \; \alpha \in (0,1).
\end{equation}
Here $a$ is the dispersion of the symmetrical random walk.}

The proof of the Proposition is based on the Fourier-Laplace transform of the characteristic function of the random walk and on the formula
$$
\hat Y^2(s) \ = \ \int \langle Y^2(t) \rangle \ e^{-st} \ dt \ = \ -\frac{\partial^2 \hat \theta(\lambda,s) }{\partial\lambda^2}|_{\lambda=0},
$$
where $ \hat \theta(\lambda, s) \ = \ \int_0^{\infty} e^{-st} \langle e^{i \lambda Y(t)} \rangle \ dt$.\\

\bigskip

The main idea of our analysis is to compare statistical characteristics of the trajectories from different families (for normal and mutant axons separately) with the mean squared displacements for CTRW given by (\ref{1})-(\ref{2}). Using the empirical law for $\langle Y^2(t) \rangle$ as a function of $t$ we can conclude which parameters of the CTRW model imply the similar law for $\langle Y^2(t) \rangle$.

\noindent On Fig. 5-10 we present our calculations of $\langle Y^2(t) \rangle$ and  $\langle (Y(t)-<Y(t)>)^2 \rangle$ using the statistical data and construct corresponding interpolation curves in each of three cases under our consideration. Here $\langle \cdot \rangle$ is the empirical average.
\\

\begin{figure}[h]
  \centering
  \includegraphics[height=7cm]{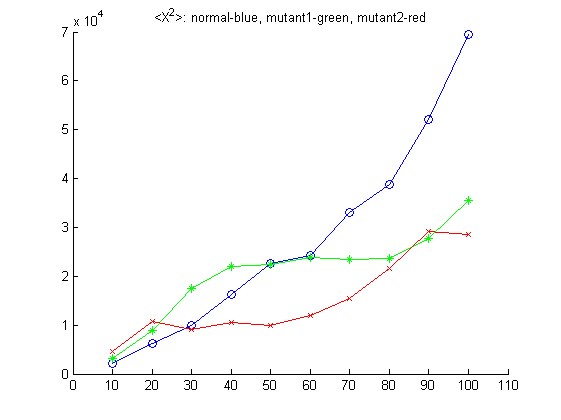}
  \caption{The first case: second moments as a function of time(\% of length)}
  \includegraphics[height=7cm]{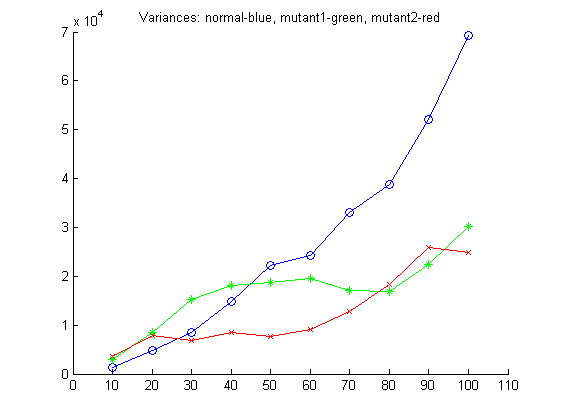}
 \caption{The first case: variances as a function of time(\% of length)}
\end{figure}

\begin{figure}[h]
  \centering
  \includegraphics[height=7cm]{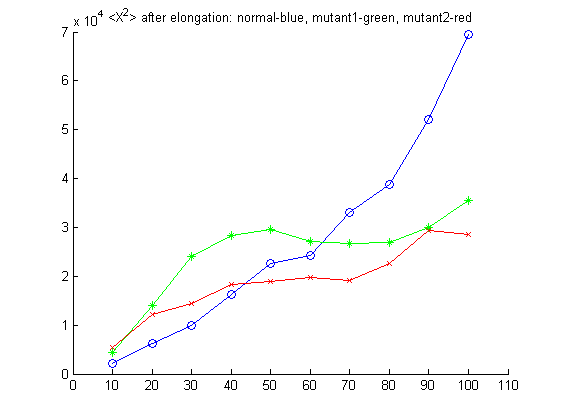}
\caption{The second case (elongation of short axons): second moments as a function of time(\% of length)}
\end{figure}

\begin{figure}[h]
\centering
  \includegraphics[height=7cm]{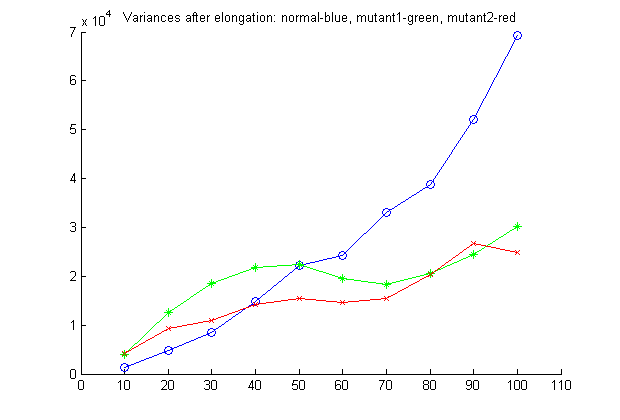}
  \caption{The second case (elongation of short axons): variances as a function of time(\% of length)}
\end{figure}

\begin{figure}[h]
  \centering
  \includegraphics[height=7cm]{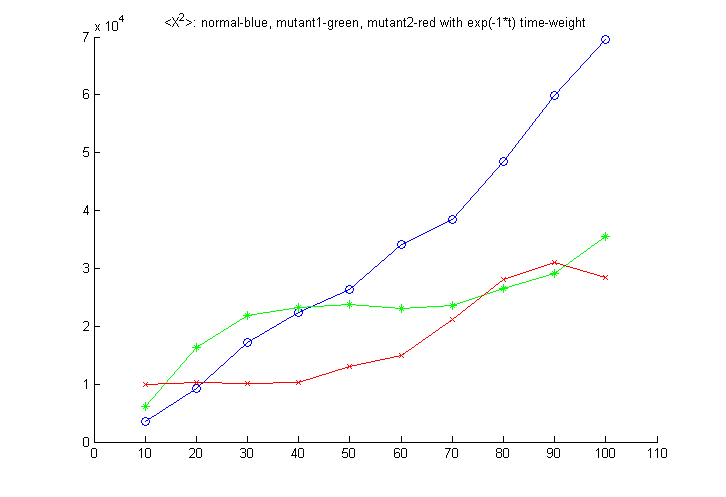}
\caption{The third case : second moments as a function of time(\% of length with weights)}
\end{figure}

\begin{figure}[h]
\centering
  \includegraphics[height=7cm]{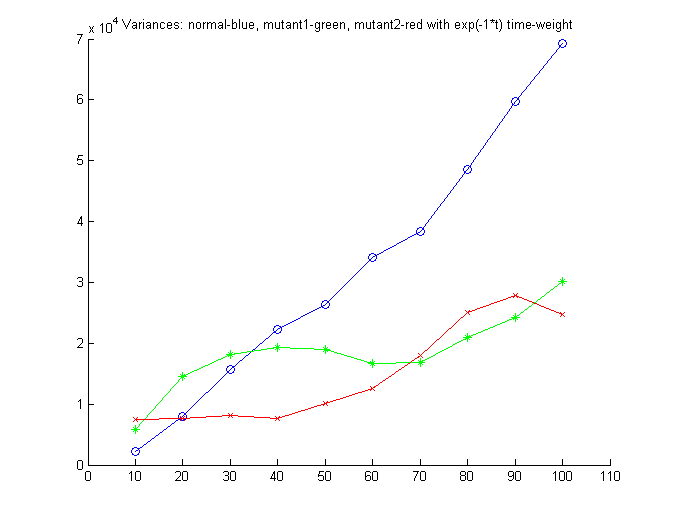}
  \caption{The third case : variances as a function of time(\% of length with weights)}
\end{figure}

\section{Conclusions}

We propose a continuous time random walk model as a model of axon growing. We can observe two different regimes during evolution of axon growth cones. The second regime, which is started approximately after $0.6 \ L$, is related with increasing coefficient $\frac{2a}{\langle \tau \rangle} $, see formula (\ref{1}). That can be the result of decreasing averaged waiting time $\langle \tau \rangle$ or increasing dispersion of the spacial random walk $a$.

The distributions of the waiting time for normal and mutant axons are different, and by the formula (\ref{1}) in the proposition the average waiting time for mutant axons is longer (greater) than for normal axons. Using our results on the shape of $
\langle Y^2(t) \rangle $ for mutant axons we can formulate a hypothesis that for normal axons the time between renewals has a finite mean, whereas for mutant axons the time between renewals is greater in average or even can have an infinite mean. In the latter case, the scaling limit for the CTRW is an operator Levy motion subordinated to the hitting time process of a classical stable subordinator, see e.g. \cite{MS}. In this case, for large enough $t$ we have $\langle Y^2(t) \rangle \ \approx  \  t^{\alpha}$ with $0< \alpha< 1$. That means that the hitting time for this CTRW to reach distant compact sets is much more greater for mutant axons than for normal ones.

\end{document}